\documentclass[11pt,draftcls,onecolumn,journal]{IEEEtran}
\IEEEoverridecommandlockouts
\usepackage{cite}
\usepackage[tbtags]{amsmath}
\usepackage{amssymb,amsfonts}
\usepackage[most]{tcolorbox}
\usepackage{enumitem}
\usepackage{graphicx}
\usepackage{textcomp}
\usepackage{xcolor}
\usepackage{caption}
\usepackage{subcaption}
\usepackage{commath}
\usepackage{url} 
\usepackage{multirow}

\definecolor{SPMBlue}{RGB}{0,82,155}
\definecolor{SPMBlueLight}{RGB}{220,232,250}

\newtcolorbox{spmalgorithm}[2][]{
    enhanced,
    colback=SPMBlueLight,
    colframe=SPMBlueLight,
    fonttitle=\bfseries,
    coltitle=black,
    title={#2},
    boxrule=0pt,
    left=0mm,
    right=0mm,
    top=0mm,
    bottom=0mm,
    arc=0mm,
    breakable,
    #1
}

\def\BibTeX{{\rm B\kern-.05em{\sc i\kern-.025em b}\kern-.08em
    T\kern-.1667em\lower.7ex\hbox{E}\kern-.125emX}}

\begin{document}
\title{Convex Optimization-Based Procedures for Non-Convex Quadratic Problems}

\author{Mahmoud Zaher, and Emil Björnson}

\maketitle

Mathematical optimization plays a fundamental role in signal processing and wireless communications, serving as an essential framework for the systematic design of modern systems. Many design challenges in these fields, as well as in many others, can naturally be formulated as optimization problems. Over the years, the advancements in signal processing applications have significantly changed the structure and complexity of these optimization problems, creating new challenges in their analysis, understanding, and solution \cite{liu2024survey}. Consequently, the rapid development of sophisticated optimization theories and algorithms tailored to the demands of next-generation systems is crucial.

Quadratic optimization problems constitute one of the most important classes of optimization problems in modern engineering systems. In signal processing and communications, quadratic forms naturally emerge when modeling power, energy, covariance matrices, and Euclidean distances, to name a few examples. Consequently, a broad family of practical design problems can be represented using quadratically constrained quadratic programs (QCQPs), where both the objective function and the constraints are quadratic functions of the optimization variables. While convex QCQPs can be solved efficiently using polynomial-time algorithms, the general non-convex QCQP remains computationally challenging. Specifically, indefinite quadratic forms and rank constraints often induce NP-hardness. Non-convex QCQP problems arise in a broad range of signal processing, communications, control, machine learning, and network optimization applications.

In the last three decades, substantial progress has been made in convex optimization-based procedures providing a principled way to construct tractable approximations and iterative algorithms capable of producing high-quality solutions for the non-convex and NP-hard QCQP problems with polynomial-time complexity. Among the most prominent methods in this regard are semidefinite relaxation (SDR) and successive convex approximation (SCA). The main idea behind SDR is to globally convexify a problem through lifting and relaxation of non-convex constraints. It has become one of the most influential approaches for handling rank-constrained formulations. On the other hand, SCA methods exploit local convexity through the construction of convex surrogate functions to approximate non-convex functions at a given point, and iteratively solve the resulting convex problems until convergence. A clear distinction between the two methods is that SDR often leads to infeasible solutions to the original problem due to relaxation, requiring post-processing of the SDR output to generate a feasible solution, whereas SCA belongs to the class of inner approximation methods, which leads to feasible solutions throughout its iterations.

Technological advancements in different fields have driven performance optimization problems to a new era characterized by a high dimension and/or a large number of constraints, which poses significant design and computational challenges. For instance, in wireless communications, the emergence of massive multiple-input multiple-output (MIMO) architectures for cellular and cell-free networks has created optimization problems involving hundreds or thousands of coupled variables and constraints \cite{shi2015large}. Solving such problems using global optimization methods is computationally infeasible. This has attracted interest in finding computationally efficient approaches for the non-convex QCQP problems. The idea is based on combining the convex optimization-based procedures mentioned earlier, with scalable first-order methods (FOMs) to enable the treatment of extremely large-scale problems.

\section*{Relevance}

Non-convex quadratic optimization has become a cornerstone of modern signal processing applications. Many engineering systems require balancing multiple objectives and constraints that naturally admit quadratic representations. Important examples in signal processing and wireless communications include beamforming design, MIMO detection, sparse signal recovery, and covariance estimation.

This lecture note article elucidates the latest advancements on solving non-convex QCQPs. We aim to provide an accessible yet rigorous overview of convex optimization-based procedures for this type of problems, from SDR and classical randomization techniques to modern iterative and large-scale optimization frameworks. The discussion culminates in multicast beamforming optimization, which serves as a representative application example where non-convex QCQPs arise naturally and where convex optimization techniques have demonstrated remarkable practical success. The goal is to bridge mathematical foundations with the underlying intuitions of various solution approaches and to highlight the interplay between convex optimization theory and modern wireless signal processing design.

\section*{Prerequisites}

The readers require basic knowledge in linear algebra and convex optimization.

\section*{Non-Convex Quadratic Problems}

In this section, we present a standard formulation for QCQPs that encompasses numerous problems that are highly relevant to the signal processing and communications community. Specifically, a general homogeneous QCQP can be formulated as
\begin{subequations}
    \begin{align}
        \mathop{\mathrm{minimize}}\limits_{\mathbf{x} \in \mathbb{C}^n}\quad &\mathbf{x}^H\mathbf{Q}\mathbf{x} \\
        \textrm{s.t.}\quad &\mathbf{x}^H\mathbf{P}_i\mathbf{x} \geq g_i, \quad i = 1, \hdots, m,
    \end{align}
    \label{QCQP}%
\end{subequations}
where $\mathbf{Q}, \mathbf{P}_1, \hdots, \mathbf{P}_m \in \mathbb{H}^n$ are general $n \times n$ Hermitian matrices.
The problem is convex if and only if the objective matrix $\mathbf{Q}$ is positive semidefinite (PSD) and all constraint matrices $\mathbf{P}_1, \hdots, \mathbf{P}_m$ are negative semidefinite. Otherwise, the problem becomes non-convex and generally NP-hard.
Note that the formulation above implicitly includes problems with equality constraints, which are equivalent to two opposing inequalities. Without loss of generality, we focus on the homogeneous QCQP \eqref{QCQP}, since a non-homogeneous QCQP, which includes an additional linear term, can be homogenized by adding an extra scalar variable; see \cite{wolkowicz2000handbook} for details.

\begin{figure}[h]
\centering
\setlength{\abovecaptionskip}{0.33cm plus 0pt minus 0pt}
\includegraphics[scale=0.5]{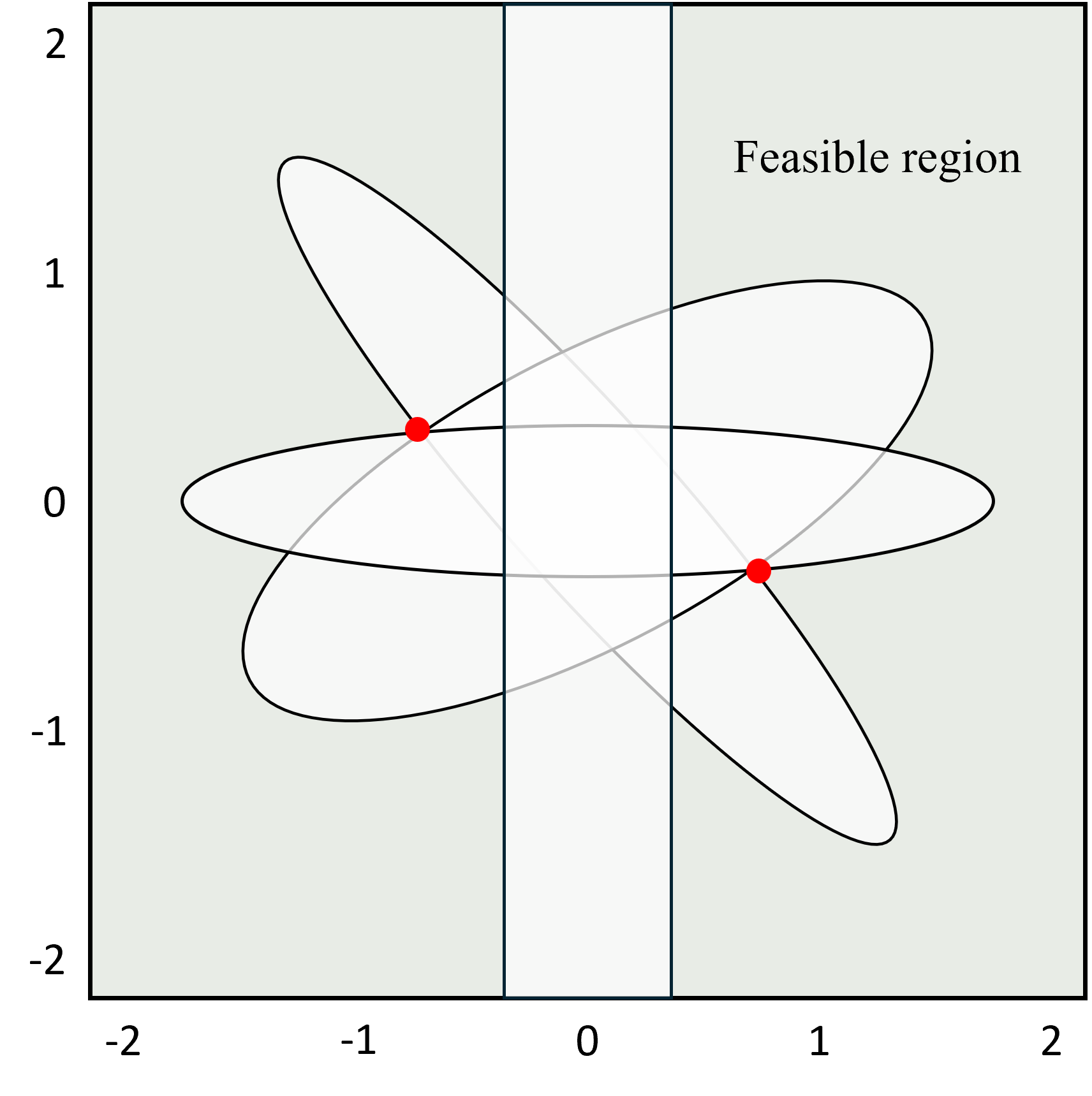}
\caption{Non-convex QCQP example in $\mathbb{R}^2$. The black lines represent the boundaries of the constraints.}
\label{QCQP_graphic}
\end{figure}

To gain some insight into the problem structure, let us consider a special instance of \eqref{QCQP}, where all variables are real-valued, $n = 2$ and all matrices $\mathbf{Q}, \mathbf{P}_1, \hdots, \mathbf{P}_m \in \mathbb{H}^n$ are positive (semi)definite. Note that this implies we are dealing with a non-convex QCQP. Figure \ref{QCQP_graphic} illustrates the intricate structure of the feasible region in this simplified scenario. As shown in the figure, the feasible region is formed by the intersection of the exteriors of multiple ellipsoids or elliptic cylinders, which makes the problem difficult even in such a simplified scenario.

\section*{Successive Convex Approximation}

SCA is one of the most widely used frameworks for solving non-convex optimization problems in modern signal processing and wireless communications. Unlike SDR, which attempts to convexify a problem globally through lifting and relaxation, SCA operates locally by iteratively constructing and solving convex approximations of the original non-convex problem. The central idea is to iteratively replace non-convex components with locally tight convex surrogates that can be solved efficiently. The resulting algorithms often exhibit lower computational complexity than classical SDR-based methods, which rely on randomization for producing a solution to the original problem, while maintaining good practical performance even for large-scale systems.

The origins of SCA can be traced to the methodology proposed by \cite{marks1978general} in the late 1970s, where non-convex constraints were replaced by conservative convex approximations that are successively updated around the current iterate. Specifically, the algorithm constructs a sequence of convex subproblems that “inner approximate” the feasible region of the original non-convex problem. Each subproblem provides a feasible solution, improving convergence and stability compared to outer approximation techniques. The method guarantees feasibility at every iteration and aims to converge to a stationary point satisfying necessary optimality conditions. The main advantages of SCA lie in its flexibility in handling complicated (non-convex) constraints and its ability to exploit problem-specific structure.

In general, a surrogate function $\tilde{f}(\mathbf{x};\mathbf{x}^{(t)})$, constructed around an arbitrary vector $\mathbf{x}^{(t)}$, for a non-convex function $f(\mathbf{x})$ used in SCA typically satisfies the following properties:
\begin{enumerate}
    \item \textbf{Local tightness:} $\tilde{f}(\mathbf{x}^{(t)};\mathbf{x}^{(t)}) = f(\mathbf{x}^{(t)})$.
    \item \textbf{First-order consistency:} $\nabla \tilde{f}(\mathbf{x}^{(t)};\mathbf{x}^{(t)}) = \nabla f(\mathbf{x}^{(t)})$.
    \item \textbf{Convexity:} $\tilde{f}(\mathbf{x};\mathbf{x}^{(t)})$ is convex in $\mathbf{x}$.
    \item \textbf{Feasibility:} either majorization or feasibility-preserving inner approximation, depending on whether the surrogate is constructed for the objective function or the constraints.
\end{enumerate}
The first two properties are primarily intended for achieving a good approximate solution in every SCA iteration. The third property ensures that the surrogate problem can be efficiently solved using accessible, usually free, optimization software. The last property ensures that the solution to the surrogate problem in every SCA iterate is also feasible for the original non-convex optimization problem. A general framework for the SCA method is provided in ``A Generic SCA Framework''.

\begin{spmalgorithm}{A Generic SCA Framework}
\textbf{Input:} Initial feasible point $\mathbf{x}^{(0)}$.

\begin{enumerate}[leftmargin=8mm,label=\arabic*.]

\item Set iteration index $t=0$.

\item Construct convex surrogate functions
$\tilde f(\mathbf{x};\mathbf{x}^{(t)})$ and $\tilde h_i(\mathbf{x};\mathbf{x}^{(t)})$.

\item Solve the convex subproblem
\[
\begin{aligned}
\mathop{\mathrm{minimize}}\limits_{\mathbf{x} \in \mathbb{C}^n} \quad &
\tilde f(\mathbf{x};\mathbf{x}^{(t)}) \\
\text{s.t.} \quad &
\tilde h_i(\mathbf{x};\mathbf{x}^{(t)}) \leq g_i,
\quad i=1,\hdots,m.
\end{aligned}
\]

\item Obtain the optimizer $\mathbf{x}^{(t+1)}$.

\item Update $t \leftarrow t+1$.

\item Repeat Steps 2-5 until convergence.

\end{enumerate}

\textbf{Output:} Stationary solution $\mathbf{x}^{\star}=\mathbf{x}^{(t)}$.

\end{spmalgorithm}

For brevity, let us consider the special instance of QCQP \eqref{QCQP}, where all matrices $\mathbf{Q}, \mathbf{P}_1, \hdots, \mathbf{P}_m \in \mathbb{H}^n$ are positive (semi)definite. Note that this specific instance of \eqref{QCQP} captures the multicast downlink beamforming optimization problem \cite{gershman2010convex}. This results in the $m$ constraints being non-convex. We highlight that if the matrices are indefinite, each constraint can then be divided into a convex and a concave part, and the same procedure will be applied for the concave parts as detailed below. The key idea of SCA is to replace the non-convex constraints at iteration $t$ by tighter convex approximations. 

The first-order Taylor approximation of a function $f(\mathbf{x}): \mathbb{C}^n \rightarrow \mathbb{R}$ around an arbitrary point $\mathbf{x}_0$ can be written as
\begin{equation}
    \tilde f(\mathbf{x}) \approx f\left(\mathbf{x}_0\right) + 2\Re\left\{\nabla_{\mathbf{x}^*}f\left(\mathbf{x}_0\right)^H\left(\mathbf{x} - \mathbf{x}_0\right)\right\}.
\end{equation}
Since $\mathbf{P}_i$ are Hermitian positive (semi)definite for all $i$, the first-order approximations of $\mathbf{x}^H\mathbf{P}_i\mathbf{x}, \forall i$, represent its global lower-bound. Accordingly, for an arbitrary vector $\mathbf{x}^{(t)}$, we have
\begin{equation}
    \mathbf{x}^H\mathbf{P}_i\mathbf{x} \geq 2\Re\left\{\left(\mathbf{x}^{(t)}\right)^H\mathbf{P}_i\mathbf{x}\right\} - \left(\mathbf{x}^{(t)}\right)^H\mathbf{P}_i\mathbf{x}^{(t)}, \quad i = 1, \hdots, m,
\end{equation}
where the right-hand side represents the first-order approximations of $\mathbf{x}^H\mathbf{P}_i\mathbf{x}, \forall i$, around $\mathbf{x}^{(t)}$. Note that the right-hand side is affine, and so convex, in $\mathbf{x}$. To that end, the following convex surrogate problem is solved at every SCA iteration:
\begin{subequations}
    \begin{align}
        \mathop{\mathrm{minimize}}\limits_{\mathbf{x} \in \mathbb{C}^n}\quad &\mathbf{x}^H\mathbf{Q}\mathbf{x} \\
        \textrm{s.t.}\quad &2\Re\left\{\left(\mathbf{x}^{(t)}\right)^H\mathbf{P}_i\mathbf{x}\right\} - \left(\mathbf{x}^{(t)}\right)^H\mathbf{P}_i\mathbf{x}^{(t)} \geq g_i, \quad i = 1, \hdots, m.
    \end{align}
    \label{SCA}%
\end{subequations}
It is easily verifiable that the first-order approximations satisfy the four conditions for the choice of the surrogate function listed earlier. Since the first-order approximations represent a lower-bound on $\mathbf{x}^H\mathbf{P}_i\mathbf{x}, \forall i$, starting from a feasible initial point at $t = 0$, the solution to the approximate convex problem produces a feasible solution to the original problem at every iteration.
After solving the surrogate convex problem, the resulting solution becomes the center point for constructing the next approximation. In particular, a sequence of the convex problem is solved such that the vector $\mathbf{x}^{(t)}$ is set to the solution of the convex problem of the previous iteration. The process repeats until convergence.

Intuitively, SCA transforms a difficult global optimization problem into a sequence of tractable local optimization problems, each of which can be solved efficiently using convex optimization techniques. Although global optimality cannot generally be guaranteed, the SCA method guarantees convergence to a local optimal solution of the original non-convex optimization problem, thanks to the four properties utilized for the choice of the surrogate functions that we mentioned earlier. The quality of the final solution heavily depends on the following factors: (i) initialization strategy, (ii) surrogate construction, (iii) problem geometry, and (iv) number of local optima.

Despite its wide adoption in many engineering applications, the SCA approach generally suffers from two limitations. The primary limitation is the lack of global optimality guarantees. Nevertheless, for many practical signal processing problems, especially large-scale problems, SCA provides one of the best trade-offs between solution quality and computational complexity. The second limitation is that it requires initialization with a feasible solution. A good initial feasible point can be challenging to obtain, particularly for large-scale problems. In fact, different initial points may result in final solutions that are of substantially different quality, which emphasizes the sensitivity of the SCA method to the chosen initialization.

\section*{Semidefinite Relaxation}

Since the beginning of the 2000s, SDR has played a major role across a wide range of applications in signal processing and communication systems \cite{wolkowicz2000handbook,luo2010semidefinite}. In general, SDR serves as a powerful and computationally efficient approximation method for solving many challenging optimization problems, including those that are potentially NP-hard. Numerous practical studies have demonstrated that SDR can achieve accurate solutions that are close to optimal. Theoretical investigations, such as those in \cite{wolkowicz2000handbook,luo2010semidefinite} and references therein, have established the approximation performance of SDR under various problem formulations. The main idea in these classical works is to combine SDR with randomization to find an approximate solution from the output high-rank solutions of SDR.

The first step to devise the SDR of \eqref{QCQP} is to recognize that
\begin{align}
\mathbf{x}^H\mathbf{Q}\mathbf{x} &= \textrm{Tr}(\mathbf{x}^H\mathbf{Q}\mathbf{x}) =  \textrm{Tr}(\mathbf{Q}\mathbf{x}\mathbf{x}^H), \\
\mathbf{x}^H\mathbf{P}_i\mathbf{x} &= \textrm{Tr}(\mathbf{x}^H\mathbf{P}_i\mathbf{x}) =  \textrm{Tr}(\mathbf{P}_i\mathbf{x}\mathbf{x}^H).
\end{align}
where $\textrm{Tr}(\cdot)$ denotes the trace of a matrix.

Specifically, SDR can be applied to the non-convex QCQP \eqref{QCQP} utilizing the lifted matrix variable $\mathbf{X} = \mathbf{x}\mathbf{x}^H$. Such formulation implies that $\mathbf{X}$ corresponds to a rank-1 Hermitian PSD matrix. 
An equivalent reformulation of \eqref{QCQP} can then be written as
\begin{subequations}
\begin{align}
\mathop{\mathrm{minimize}}\limits_{\mathbf{X}} \quad & \textrm{Tr}(\mathbf{Q} \mathbf{X}) \\
\text{s.t.} \quad & \textrm{Tr}(\mathbf{P}_i \mathbf{X}) \geq g_i, \quad i = 1, \hdots, m,
\\
& \mathbf{X} \succeq 0, \quad \text{rank}(\mathbf{X}) = 1,
\end{align}
\label{SDP}%
\end{subequations}
where $\mathbf{X} \succeq 0$ is used to indicate that $\mathbf{X}$ is PSD.

The reformulated problem \eqref{SDP} is still non-convex, and it is as difficult to solve as the QCQP \eqref{QCQP}. The primary difference in \eqref{SDP} is that the non-convexity is now confined in the rank constraint, $\text{rank}(\mathbf{X}) = 1$, which reveals the fundamental difficulty in solving \eqref{QCQP}. The objective and all other constraints are now linear, and so convex, in $\mathbf{X}$. The idea of SDR is now clear---drop the non-convex rank constraint to obtain the following relaxed version of \eqref{SDP}:
\begin{subequations}
\begin{align}
\mathop{\mathrm{minimize}}\limits_{\mathbf{X}} \quad & \textrm{Tr}(\mathbf{Q} \mathbf{X}) \\
\text{s.t.} \quad & \textrm{Tr}(\mathbf{P}_i \mathbf{X}) \geq g_i, \quad i = 1, \hdots, m,
\\
& \mathbf{X} \succeq 0.
\end{align}
\label{SDR}%
\end{subequations}
Problem \eqref{SDR} is a standard semidefinite programming (SDP) problem. The key advantage of the formulation in \eqref{SDR} is that it can be solved with arbitrary accuracy in a numerically stable and computationally efficient manner. Moreover, SDR problems can be addressed conveniently and effectively using widely available free software packages; see \cite{luo2010semidefinite} for an example.

After solving the relaxed problem, the only immediate conclusion that can be drawn is that the obtained objective value serves as a lower bound on the optimal value of the original NP-hard problem, since ignoring the rank constraint expands the feasible set. If the optimal solution to \eqref{SDR}, denoted as $\mathbf{X}^\star$, is in fact rank-1, then our relaxation was not really a relaxation in the sense that the optimal solution $\mathbf{X}^\star$ is also a solution to \eqref{SDP}. In this case, we can directly obtain the global optimal solution to \eqref{QCQP} exactly, denoted as $\mathbf{x}^\star$, by simply extracting the principal component of $\mathbf{X}^\star$. Obviously, there must be a catch to this approach, for otherwise we would have solved an NP-hard problem in polynomial time. It turns out that standard interior-point SDP solvers will always favor a higher-rank solution over its lower-rank counterpart, even if both can achieve the same objective value. The reason is that low-rank solutions to SDP problems possess numerous zero eigenvalues and therefore tend to reside on the boundary of the feasible set rather than in its interior. This can be easily verified as a small additive perturbation to these zero eigenvalues can render the solution infeasible by violating the PSD constraint on the solution matrix. This results in achieving the solution with maximum rank among all possible optimum solutions, requiring post-processing of the SDP solution matrix to produce an approximate rank-1 solution.

Besides the rank-relaxation interpretation of SDR, an alternative interpretation can be derived from Lagrangian duality. Specifically, the SDR formulation in \eqref{SDR} can be shown to correspond to the Lagrangian bidual of the original problem in \eqref{QCQP}. As a result, it represents the convex hull, or tightest convex relaxation, of the original optimization problem.

\subsection*{Randomization Methods}

Researchers in optimization have developed techniques to extract an approximate solution $\hat{\mathbf{x}}$ for the original QCQP \eqref{QCQP} from the high-rank solution $\mathbf{X}^\star$ to the SDP \eqref{SDR}. The first classical idea is based on randomization; that is, using $\mathbf{X}^\star$ to generate many candidate vectors and selecting the best candidate that provides the lowest objective value. Several randomization methods have been proposed in the literature. In the following, we briefly describe some of these methods, which are also presented in \cite{sidiropoulos2006transmit}. 

Let the eigen decomposition of the Hermitian PSD solution matrix be defined as $\mathbf{X}^\star = \mathbf{U}\boldsymbol{\Sigma} \mathbf{U}^H$. The candidate vectors can then be generated according to $\mathbf{x}_n = \mathbf{U}\boldsymbol{\Sigma}^\frac{1}{2}\mathbf{e}_n$, $n = 1, \hdots, N$. The elements of the random vectors $\mathbf{e}_n$ can be chosen to be independent random variables that are uniformly distributed on the unit circle in the complex plane; i.e., $[\mathbf{e}_n]_i = e^{j\theta_{n,i}}$, where the angles $\theta_{n,i}, \forall n, i$ are independent and uniformly distributed in the range $[0, 2\pi)$. This ensures that $\mathbf{x}_n^H\mathbf{x}_n = \textrm{Tr}(\mathbf{X}^\star)$, regardless of the specific realization of $\mathbf{e}_n$. Another option is to choose $\mathbf{e}_n \sim \mathcal{CN}(\mathbf{0}, \mathbf{I}_N)$, which denotes a zero-mean unit-variance circularly symmetric complex Gaussian random variable. In both cases, the following holds true for the randomized vectors: $\mathbb{E}\{\mathbf{x}\mathbf{x}^H\} = \mathbf{X}^\star$, which entails $\mathbb{E}\{\mathbf{x}^H\mathbf{Q}\mathbf{x}\} = \textrm{Tr}(\mathbf{QX^\star})$. This means that, on average, the objective value of the randomized vectors is equal to the SDR lower bound. For both uniform and Gaussian randomization, each randomized vector needs to be scaled to satisfy the constraints, and the best candidate is retained. Although randomization does not guarantee any kind of optimality, it often yields reasonable approximations for relatively small-scale problems. An overview of the results on the approximation accuracy of randomization methods can be found in \cite{luo2010semidefinite}.

The main problem with randomization is that an increasingly large number of candidate solutions must be tested to find a satisfactory approximation when the problem size grows, which results in high computational requirements. This necessitates the development of an efficient procedure to extract a near-optimal rank-1 solution from the higher-rank solution obtained via SDR.

\subsection*{Successive Elimination Algorithm}

As discussed earlier, when utilizing SDR to solve the QCQP \eqref{QCQP}, the resulting solution matrix will most likely have high rank. In many QCQP formulations, SDR followed by randomization provides a ``good'' solution but may still exhibit large performance gaps to the global optimum due to the mismatch between the relaxed SDP \eqref{SDR} and the original rank-constrained problem \eqref{SDP}. In particular, randomization methods often provide satisfactory performance for small-scale problems, however, the technological advancements in different engineering fields have driven the need for solving much larger-scale optimization problems. The development of more efficient and scalable procedures for solving this type of problem is thus vital. To this end, a novel iterative rank reduction procedure called \emph{Successive Elimination Algorithm (SEA)} was proposed in \cite{zaher2026cell} to extract a near-globally optimal rank-1 solution to \eqref{SDP}.

This idea can be viewed as a generalization of the widespread approach of penalty methods, where a relaxation of a hard problem is performed to make the problem more tractable, then a penalty is introduced to tighten the optimality gap introduced by the relaxation (cf. common description of penalty methods as detailed in \cite{liu2024survey}). In our case, any formulation that attempts to directly reduce the rank of the solution matrix produced by SDR will result in a non-convex problem that cannot be handled in a straightforward manner, and hence the need for the iterative penalty procedure.

Specifically, at each iteration of the SEA, a penalty term in the form of a quadratic matrix product is added to the objective function. As such, the following penalized problem is solved at each iteration until a rank-1 solution is achieved:
\begin{subequations}
\begin{align}
\mathop{\mathrm{minimize}}\limits_{\mathbf{X}} \quad & \textrm{Tr}(\mathbf{Q} \mathbf{X}) + \zeta\sum_j\mathbf{u}_j^H\mathbf{X}\mathbf{u}_j \\
\text{s.t.} \quad & \textrm{Tr}(\mathbf{P}_i \mathbf{X}) \geq g_i, \quad i = 1, \hdots, m,
\\
& \mathbf{X} \succeq 0,
\end{align}
\label{SEA}%
\end{subequations}
The vectors $\mathbf{u}_j, \forall j$, appearing in the additive penalty terms are chosen as the eigenvectors corresponding to the second-largest eigenvalues of the higher-rank solutions obtained in previous iterations, whereas $\zeta$ is a constant penalty factor that is chosen to balance between the original objective and the introduced penalties. The goal of these penalty terms is to push the solver away from the high-rank solutions produced by solving a sequence of the (penalized) relaxed problem. Because the relaxed SDP solution matrices are Hermitian PSD and possess distinct nonzero eigenvalues, their eigenvectors are mutually orthogonal. Consequently, incorporating these eigenvectors into the penalty terms has only a limited impact on the achieved objective value of the original optimization problem \eqref{SDP}. Furthermore, the eigenvectors associated with the second-largest eigenvalues cannot yield an optimal rank-1 solution, since the dominant eigenvector of the same solution always yields a superior objective value. Therefore, penalizing these secondary eigenvectors minimally affects subsequent iterations while progressively steering the solution toward a near-globally optimal rank-1 solution. The idea of the elimination procedure is summarized in ``SEA Framework''.

\begin{spmalgorithm}{SEA Framework}

\begin{enumerate}[leftmargin=8mm,label=\arabic*.]

\item Set iteration index $j = 0$.

\item Solve the convex SDP problem \eqref{SDR} globally.

\item Obtain the optimizer $\mathbf{X}^{(j)}$.
    
\item \textbf{If} $\mathbf{X}^{(j)}$ is not rank-1

\begin{enumerate}[leftmargin=4mm,label=\alph*.]

\item Set $\mathbf{u}_j \leftarrow$ Eigenvector corresponding to the second-largest eigenvalue of $\mathbf{X}^{(j)}$.

\item Update $j \leftarrow j+1$.
    
\item Update $\mathbf{X}^{(j)} \leftarrow$ Solution to the penalized convex problem \eqref{SEA}.

\item Repeat Steps a-c until a rank-1 solution is achieved.

\end{enumerate}

\item If $\mathbf{X}^{(j)}$ is rank-1, its principal component is the optimal solution to QCQP \eqref{QCQP}.

\end{enumerate}

\end{spmalgorithm}

An advantage of the SEA is that it does not require an initialization with a feasible solution, which is necessary for algorithms that rely on the SCA method. In many cases, finding an effective feasible solution for initialization is a difficult problem on its own, particularly for large-scale problems. A theoretical proof of the approximation performance of the SEA is currently lacking; nevertheless, it has demonstrated superior practical performance compared to randomization methods and SCA-based solutions \cite{zaher2026cell}.

\section*{Implementation Aspects}

Modern convex optimization-based procedures for non-convex QCQPs consist of two distinct layers: a high-level convexification strategy and a low-level numerical optimization solver. The former transforms the original non-convex problem into one or more convex subproblems, whereas the latter computes the numerical solution of these subproblems. This distinction is particularly important because the same convexification procedure may be implemented using different optimization solvers depending on the problem size, desired accuracy, and computational resources.

\begin{figure}[h]
\centering
\setlength{\abovecaptionskip}{0.33cm plus 0pt minus 0pt}
\includegraphics[scale=0.2]{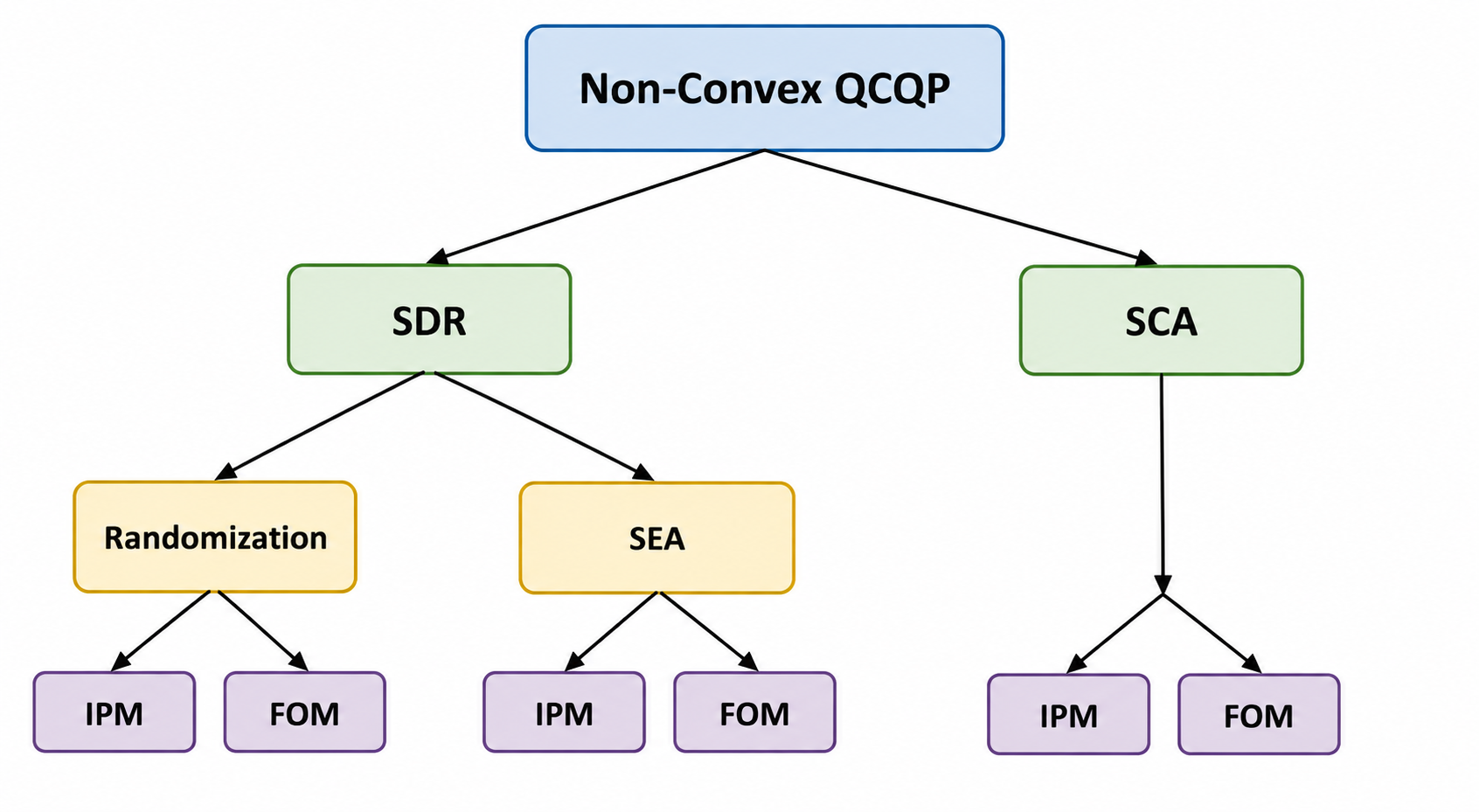}
\caption{Convex optimization procedures and numerical solvers for non-convex QCQPs.}
\label{tree_soln_app}
\end{figure}

Historically, interior-point methods (IPMs) have served as the workhorse for solving convex optimization problems due to their strong convergence guarantees and high numerical accuracy. However, the emergence of large-scale problems has motivated increasing interest in FOMs, which often provide substantially lower computational complexity with moderate numerical accuracy. Fig. \ref{tree_soln_app} illustrates the different layers of the convex optimization-based procedures for solving non-convex QCQPs.

For SDP problems, IPMs often achieve polynomial-time complexity and high numerical precision. Specifically, the worst-case computational complexity for solving an SDP involving an $n \times n$ matrix variable for a generic IPM solver is $\mathcal{O}(n^{6.5})$. However, the average complexity can be far lower than this worst-case bound with typical complexities of $\mathcal{O}(n^{3.5})$ to $\mathcal{O}(n^{4.5})$ depending on the problem structure; see \cite{zaher2026cell} and references therein. As for IPMs employed for solving the convex surrogate problem in the SCA method, the typical complexity per SCA iteration is $\mathcal{O}((n+m)^{3.5})$, which is comparable to that of solving the SDP. In addition, it is worth mentioning that, on average, SEA requires less iterations to converge to its final solution than SCA does. The above complexities illustrate why large-scale optimization problems quickly become computationally prohibitive to solve with IPM solvers, whether for SEA or SCA.

There are various FOMs that can be utilized to solve the convex problems associated with SEA and SCA. Many practical studies show that the alternating direction method-of-multipliers (ADMM) algorithm provides superior performance compared to other FOMs for large-scale optimization problems in the majority of simulation settings \cite{konar2017fast}. It has thus been one of the most widespread FOMs for the implementation of large-scale optimization algorithms. For that reason, we will focus our discussion about FOMs to the ADMM algorithm. The complexity involved with the implementation of ADMM for solving SEA and SCA problems heavily depend on the specific problem at hand. Representative examples of SEA-ADMM and SCA-ADMM implementations for solving non-convex QCQPs in \cite{chen2017admm,zaher2025low}, demonstrate a total computational complexity of order $\mathcal{O}(n^3)$ in both cases. Table \ref{tab:comparison} gives a high-level comparison of the convex optimization procedures and numerical solvers presented in this lecture note.

\begin{table*}[h]
\centering
\caption{Comparison of convex optimization procedures and numerical solvers.}
\label{tab:comparison}
\renewcommand{\arraystretch}{1.3}
\begin{tabular}{|l|l|l|l|l|}
\hline
\textbf{Procedure} & \textbf{Typical Convex Subproblem} & \textbf{Solver} & \textbf{Complexity} & \textbf{Solution Quality} \\
\hline

SDR + Randomization & SDP & IPM & High & Moderate \\
\hline

SDR + Randomization & SDP & ADMM & Medium & Moderate \\
\hline

SEA & SDP & IPM & Medium & Near-global \\
\hline

SEA & SDP & ADMM & Low & Near-global \\
\hline

SCA & Quadratic & IPM & Medium & Local optimum \\
\hline

SCA & Quadratic & ADMM & Low & Local optimum \\
\hline

\end{tabular}
\end{table*}

\section*{Application Example: Multicast Beamforming Optimization}

Physical-layer multicasting constitutes one of the most prominent applications of non-convex quadratic optimization in wireless communications \cite{sidiropoulos2006transmit,gershman2010convex}. It enables efficient transmission by leveraging beamforming at the transmitters and the broadcast nature of wireless channels to deliver the same data required by multiple user equipments (UEs) using shared transmission and time-frequency resources. It serves as a vital ingredient in modern wireless networks since it can support various applications such as video streaming of live events to multiple subscribers over mobile networks, virtual reality, internet of things (IoT) applications, videoconferencing, and broadcasting federated learning models, where large common information needs to be communicated to multiple nodes simultaneously. One key benefit of multicasting is to omit unnecessary co-channel interference among UEs requesting the same information. However, a major challenge lies in determining an appropriate beamforming design that ensures a satisfactory minimum data rate for all UEs within the multicast group, since the system performance is limited by the weakest UE \cite{zaher2026cell}. 

Unlike conventional unicast transmission, where each UE receives an independent data stream and has a dedicated beamforming vector, multicast beamforming must boost the desired signal of all UEs in a multicast group using a single beamforming vector, which naturally leads to challenging non-convex optimization formulations. As a result, the effectiveness of multicast beamforming depends much on how well one can handle non-convex QCQPs.

Consider a multicast communication system consisting of a single transmitter equipped with $N$ antennas and $K$ single-antenna UEs randomly located within a large coverage area. All UEs are served simultaneously through a common multicast transmission. A narrow-band channel model is assumed, whereby for each channel realization the channel remains frequency-flat and quasi-static in time. Furthermore, perfect channel state information is assumed to be available everywhere. The channel from the transmitter to UE $k$, normalized by the noise power, is represented by $\mathbf{h}_{k} \in \mathbb{C}^N$. Accordingly, the received signal at UE $k$ can be expressed as
\begin{equation}
    y_{k}^\mathrm{dl} = \mathbf{h}_{k}^H\mathbf{w}s +  n_{k},
\end{equation}
where $s$ represents the zero-mean unit-variance multicast signal intended for all UEs, $\mathbf{w} \in \mathbb{C}^N$ is the common multicast beamforming vector, and $n_{k} \sim \mathcal{N}_{\mathbb{C}}(0, 1)$ denotes the normalized Gaussian noise at UE $k$. The signal-to-noise-ratio (SNR) of UE $k$ is thus given by $\textrm{SNR}_{k}^{\mathrm{dl}} = |\mathbf{h}_{k}^H\mathbf{w}|^2$. Since all UEs in the multicast group need to be able to decode the data, the minimum SNR across the group determines what rate can be utilized for the multicast transmission. Consequently, multicast beamforming optimization generally focuses on the quality-of-service (QoS) and max-min fair (MMF) objectives. 

For the QoS problem, the objective is to minimize the transmit power while guaranteeing a predefined target SNR for all multicast UEs, leading to the following formulation:
\begin{subequations}
    \begin{align}
        \mathop{\mathrm{minimize}}\limits_{\mathbf{w}}\quad &\norm{\mathbf{w}}^2 \\
        \textrm{s.t.}\quad &|\mathbf{h}_k^H\mathbf{w}|^2 \geq \gamma, \quad k = 1, \hdots, K,\label{SNR_constraints}
    \end{align}
    \label{QoS_problem}%
\end{subequations}
where $\gamma$ represents the SNR constraint for all UEs receiving the multicast transmission. Since $|\mathbf{h}_k^H\mathbf{w}|^2 = \mathbf{w}^H\mathbf{H}_k\mathbf{w} \geq \gamma$, with $\mathbf{H}_k = \mathbf{h}_k\mathbf{h}_k^H$, defines the exterior of an ellipsoid or elliptic cylinder, the feasible region is non-convex. This results in a non-convex QCQP problem.

For the MMF problem, the objective is to maximize the minimum SNR among the multicast UEs subject to a total power constraint at the transmitter, which can be viewed as the inverse of the QoS problem. Hence, the MMF multicast problem can be formulated as follows:
\begin{subequations}
    \begin{align}
        \mathop{\mathrm{maximize}}\limits_{\mathbf{w}} \min_{k \in \{1, \hdots, K\}} \quad &|\mathbf{h}_k^H\mathbf{w}|^2 \\
        \textrm{s.t.}\qquad &\norm{\mathbf{w}}^2 \leq P_\mathrm{max},
    \end{align}
\end{subequations}
where $P_{\mathrm{max}}$ defines the maximum transmitter power. Likewise, the MMF multicast problem is a non-convex QCQP problem due to the non-convex objective function.

\subsection*{SCA-Based Solution}

The first-order approximation of the SNR constraints in \eqref{SNR_constraints} is given by the following affine constraints in $\mathbf{w}$ at a given SCA iteration $t$:
\begin{equation}
    2\Re\left\{\left(\mathbf{w}^{(t)}\right)^H\mathbf{H}_k\mathbf{w}\right\} - \left(\mathbf{w}^{(t)}\right)^H\mathbf{H}_k\mathbf{w}^{(t)} \geq \gamma, \quad k = 1, \hdots, K.
\end{equation}
Hence, an approximation of the QoS problem in \eqref{QoS_problem} at iteration $t$ can be formulated as
\begin{subequations}
    \begin{align}
        \mathop{\mathrm{minimize}}\limits_{\mathbf{w}}\quad &\norm{\mathbf{w}}^2 \\
        \textrm{s.t.}\quad &2\Re\left\{\left(\mathbf{w}^{(t)}\right)^H\mathbf{H}_k\mathbf{w}\right\} - \left(\mathbf{w}^{(t)}\right)^H\mathbf{H}_k\mathbf{w}^{(t)} \geq \gamma, \quad k = 1, \hdots, K,
    \end{align}
\end{subequations}
which is a convex quadratic optimization problem with affine constraints in $\mathbf{w}$. As previously stated, starting from an initial feasible solution, a sequence of the convex problem is solved such that the vector $\mathbf{w}^{(t)}$ is set to the solution of the convex surrogate problem of the previous iteration. The procedure is repeated until convergence to a stationary point of the original non-convex QoS problem.

Utilizing the same approach, the convex surrogate problem at iteration $t$ for the MMF multicast problem is given by
\begin{subequations}
    \begin{align}
        \mathop{\mathrm{maximize}}\limits_{\mathbf{w}} \min_{k \in \{1, \hdots, K\}} \quad &2\Re\left\{\left(\mathbf{w}^{(t)}\right)^H\mathbf{H}_k\mathbf{w}\right\} - \left(\mathbf{w}^{(t)}\right)^H\mathbf{H}_k\mathbf{w}^{(t)} \\
        \textrm{s.t.}\qquad &\norm{\mathbf{w}}^2 \leq P_\mathrm{max}.
    \end{align}
\end{subequations}
To solve the MMF convex approximation problem, an auxiliary variable $z$ is introduced to lower-bound the SNRs of the UEs. The resulting equivalent problem is formulated as
\begin{subequations}
    \begin{align}
        \mathop{\mathrm{maximize}}\limits_{\mathbf{w}, z} \quad &z \\
        \textrm{s.t.}\quad &2\Re\left\{\left(\mathbf{w}^{(t)}\right)^H\mathbf{H}_k\mathbf{w}\right\} - \left(\mathbf{w}^{(t)}\right)^H\mathbf{H}_k\mathbf{w}^{(t)} \geq z, \quad k = 1, \hdots, K, \\
        &\norm{\mathbf{w}}^2 \leq P_\mathrm{max}.
    \end{align}
\end{subequations}
This problem is convex since the objective is linear in $z$, the approximate SNR constraints are affine in $\{\mathbf{w}, z\}$ and the power constraint is convex in $\mathbf{w}$.
Therefore, each SCA iteration can be solved efficiently using standard convex optimization software. The resulting algorithm solves a sequence of convex approximation problems monotonically improving the minimum UE SNR and converging to a stationary point of the original non-convex MMF problem. The extension of the SCA method to solve the multigroup multicast problem, where multiple multicast groups each having a dedicated beamforming vector are served on the same time-frequency resources simultaneously, can be found in \cite{hsu2017joint}.

\subsection*{SDR-Based Solution}

To solve the QoS and MMF multicast problems using SDR, we define the lifted matrix variables $\mathbf{W} = \mathbf{ww}^H$ and $\mathbf{H}_k = \mathbf{h}_k\mathbf{h}_k^H$. The multicast QoS problem can then be equivalently reformulated as
\begin{subequations}
    \begin{align}
        \mathop{\mathrm{minimize}}\limits_{\mathbf{w}} \quad &\textrm{Tr}\left(\mathbf{W}\right) \\
        \textrm{s.t.}\quad&\textrm{Tr}\left(\mathbf{H}_k\mathbf{W}\right) \geq \gamma,\quad  k = 1, \hdots, K,\\
    &\mathbf{W} \succeq \mathbf{0}, \quad \textrm{rank}\left(\mathbf{W}\right) = 1.\label{rank1}
    \end{align}
\end{subequations}
The only source of non-convexity is the rank-1 constraint. After dropping the non-convex rank constraint, the relaxed convex SDP problem can be solved using a standard solver.

As for the MMF multicast problem, a similar SDP formulation can be derived using the auxiliary variable $z$:
\begin{subequations}
    \begin{align}
        \mathop{\mathrm{maximize}}\limits_{\mathbf{w}, z}  \quad &z \\
        \textrm{s.t.}\quad &\textrm{Tr}\left(\mathbf{H}_k\mathbf{W}\right) \geq z, \quad k = 1, \hdots, K, \\
        &\textrm{Tr}\left(\mathbf{W}\right) \leq P_\mathrm{max}, \\
        &\mathbf{W} \succeq \mathbf{0}, \quad \textrm{rank}\left(\mathbf{W}\right) = 1\label{rank1}.
    \end{align}
\end{subequations}
Likewise, the non-convex rank-1 constraint is dropped to obtain the relaxed convex SDP problem that can be solved using a standard SDP solver.

For both the QoS and MMF multicast problems, if the resulting solution is rank-1, the globally optimal multicast beamformer can be recovered directly. Unfortunately, the resulting solution matrix $\mathbf{W}$ will most likely have high rank that violates the rank constraint of the original problem. Post processing of the solution to the SDP problem is thus essential. The randomization methods and SEA presented earlier can be utilized to obtain an approximate rank-1 solution. Extension to the multigroup multicast problem and further details about the randomization methods and elimination procedure in this case can be found in \cite{zaher2026cell}.

\subsection*{Numerical Examples}

In this section, numerical simulation examples for the QoS and MMF multicast beamforming problems are provided to showcase the performance of the different optimization procedures and numerical solvers presented in this lecture note. We consider that $K = 30$ UEs are uniformly distributed in an area of $750\,\textrm{m} \times 750\,\textrm{m}$. To test the presented solution approaches under different simulation settings, two simulation scenarios are considered with either distributed or co-located $N = 36$ transmitter antennas.

\begin{figure}
\setlength{\abovecaptionskip}{0.3cm plus 0pt minus 0pt}
\begin{subfigure}{.49\textwidth}
\flushleft
\includegraphics[scale=0.44]{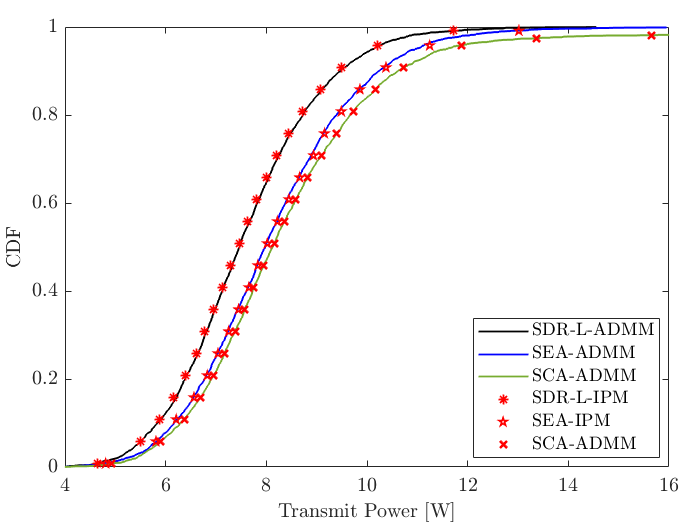}
\caption{QoS problem.}
\label{QoS_powers_fig}
\end{subfigure}
\begin{subfigure}{.49\textwidth}
\flushright
\includegraphics[scale=0.44]{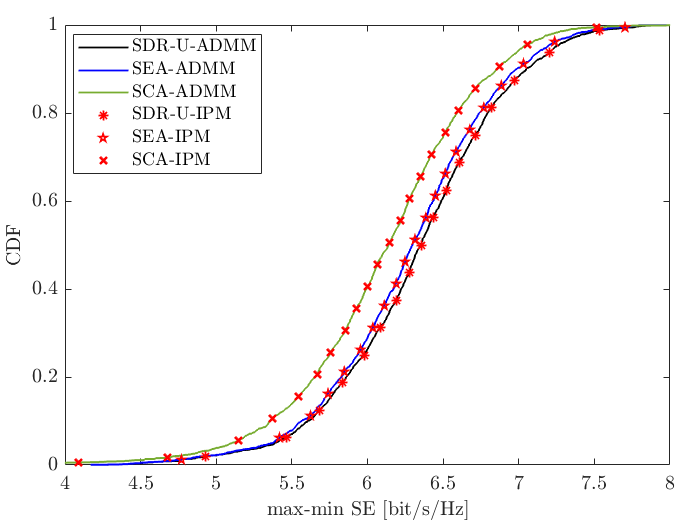}
\caption{MMF problem.}
\label{max_min_SE_fig}
\end{subfigure}
\caption{Performance of different convex optimization procedures and numerical solvers for $K = 30$ UEs and $N = 36$, (a) distributed, and (b) co-located, antennas.}
\label{results_fig}
\end{figure}

Fig. \ref{results_fig} plots the achievable performance of the different convex optimization procedures and numerical solvers for (a) required transmit power of the QoS problem with distributed transmitter antennas, and (b) max-min spectral efficiency (SE) of the MMF problem with co-located transmitter antennas. The high-rank SDR lower (or upper) bound for the QoS (or MMF) problem is plotted for comparison. It is clear that the ADMM-based solvers can achieve roughly the same performance as the IPM-based solvers for all simulation settings, regardless of the choice of the optimization procedure. This shows that the solution accuracy of the ADMM-based solvers is sufficient for the multicast beamforming problem. We highlight that ADMM offers around an order-of-magnitude lower computational time than its IPM counterpart for the same optimization procedure.

Moreover, Fig. \ref{QoS_powers_fig} shows that SEA can achieve a small reduction in the required transmit power over SCA for the same SNR target. Additionally, Fig. \ref{max_min_SE_fig} shows a more significant max-min SE improvement for SEA over SCA, with SEA reaching a small gap to the SDR upper bound. It is also worth noting that SCA failed to produce a local optimal solution for about $2\,\%$ of the samples in these setups. The reason lies in the difficulty of finding an effective feasible initial point, which is necessary for SCA-based algorithms, unlike SEA, which does not require an initial feasible solution. Note that this issue can become more severe as the problem size grows larger. We point out that the classical approach of SDR followed by randomization fails to compete with the more advanced SEA and SCA algorithms, offering degraded performance with higher computational requirements.

Finally, we would like to emphasize that multicast beamforming serves not only as an important application in its own right but also as a representative case study demonstrating how convex optimization-based procedures can effectively address challenging non-convex quadratic optimization problems.

\section*{Lessons Learned}

Non-convex quadratic optimization remains a vibrant and rapidly evolving research area for diverse engineering domains. Convex optimization-based procedures, such as SCA and SEA, have fundamentally reshaped the way that these computationally difficult problems are approached. In SCA, a feasibility-preserving inner approximation strategy is employed to construct convex surrogate constraints, and then iteratively solve a sequence of convex approximate problems until convergence to a stationary point of the original non-convex problem. As for SEA, an outer approximation strategy is employed, where penalty terms are iteratively added to tighten the relaxation gap introduced by SDR while having minimal effect on the achievable objective value.
The presented optimization procedures have demonstrated that near-optimal approximations with scalable polynomial-time algorithms are attainable for this class of non-convex and NP-hard problems. The combination of these procedures with fast FOMs provides practically good solution accuracy with ultra-low complexity implementations that is adequate for many engineering applications. Future developments will likely continue integrating optimization theory with emerging machine learning techniques and distributed computational architectures.

\section*{Acknowledgments}

This work was supported by the SUCCESS grant from the Swedish Foundation for Strategic Research.

\section*{Authors}

Mahmoud Zaher (mahmoudz@kth.se) is a postdoctoral researcher at the Department of Communication Systems, KTH Royal Institute of Technology, Sweden.

Emil Björnson (emilbjo@kth.se) is a professor of wireless communications and the head of the Department of Communication Systems, KTH Royal Institute of Technology, Sweden.

\bibliographystyle{ieeetr}
\bibliography{papercites}

\begin{thebibliography}{10}

\bibitem{liu2024survey}
Y.-F. Liu, T.-H. Chang, M.~Hong, Z.~Wu, A.~M.-C. So, E.~A. Jorswieck, and W.~Yu, ``A survey of recent advances in optimization methods for wireless communications,'' {\em IEEE Journal on Selected Areas in Communications}, vol.~42, no.~11, pp.~2992--3031, 2024.

\bibitem{shi2015large}
Y.~Shi, J.~Zhang, B.~O'Donoghue, and K.~B. Letaief, ``Large-scale convex optimization for dense wireless cooperative networks,'' {\em IEEE Transactions on Signal Processing}, vol.~63, no.~18, pp.~4729--4743, 2015.

\bibitem{wolkowicz2000handbook}
H.~Wolkowicz, R.~Saigal, and L.~Vandenberghe, {\em Handbook of semidefinite programming: theory, algorithms, and applications}.
\newblock Springer Science \& Business Media, 2000.

\bibitem{marks1978general}
B.~R. Marks and G.~P. Wright, ``A general inner approximation algorithm for nonconvex mathematical programs,'' {\em Operations research}, vol.~26, no.~4, pp.~681--683, 1978.

\bibitem{gershman2010convex}
A.~B. Gershman, N.~D. Sidiropoulos, S.~Shahbazpanahi, M.~Bengtsson, and B.~Ottersten, ``Convex optimization-based beamforming,'' {\em IEEE Signal Processing Magazine}, vol.~27, no.~3, pp.~62--75, 2010.

\bibitem{luo2010semidefinite}
Z.-Q. Luo, W.-K. Ma, A.~M.-C. So, Y.~Ye, and S.~Zhang, ``Semidefinite relaxation of quadratic optimization problems,'' {\em IEEE Signal Processing Magazine}, vol.~27, no.~3, pp.~20--34, 2010.

\bibitem{sidiropoulos2006transmit}
N.~D. Sidiropoulos, T.~N. Davidson, and Z.-Q. Luo, ``Transmit beamforming for physical-layer multicasting,'' {\em IEEE transactions on signal processing}, vol.~54, no.~6, pp.~2239--2251, 2006.

\bibitem{zaher2026cell}
M.~Zaher, E.~Bj{\"o}rnson, and M.~Petrova, ``Cell-free beamforming design for physical layer multigroup multicasting,'' {\em IEEE Transactions on Wireless Communications}, vol.~25, pp.~5262--5274, 2026.

\bibitem{konar2017fast}
A.~Konar and N.~D. Sidiropoulos, ``Fast approximation algorithms for a class of non-convex {QCQP} problems using first-order methods,'' {\em IEEE Transactions on Signal Processing}, vol.~65, no.~13, pp.~3494--3509, 2017.

\bibitem{chen2017admm}
E.~Chen and M.~Tao, ``{ADMM}-based fast algorithm for multi-group multicast beamforming in large-scale wireless systems,'' {\em IEEE Transactions on Communications}, vol.~65, no.~6, pp.~2685--2698, 2017.

\bibitem{zaher2025low}
M.~Zaher and E.~Bj{\"o}rnson, ``Low-complexity {ADMM}-based multicast beamforming in cell-free massive {MIMO} systems,'' {\em arXiv preprint [Online]. Available: https://arxiv.org/abs/2511.07310}, 2025.

\bibitem{hsu2017joint}
G.-W. Hsu, B.~Liu, H.-H. Wang, and H.-J. Su, ``Joint beamforming for multicell multigroup multicast with per-cell power constraints,'' {\em IEEE Transactions on Vehicular Technology}, vol.~66, no.~5, pp.~4044--4058, 2017.

\end{thebibliography}

\end{document}